# Properties and characterization of ALD grown dielectric oxides for MIS structures

S. Gierałtowska[1*], D. Sztenkiel[1], E. Guziewicz[1], M. Godlewski[1,2], G. Łuka[1], B.S. Witkowski[1], Ł. Wachnicki[1], E. Łusakowska[1], T. Dietl[1,3] and M. Sawicki[1]

[1] Institute of Physics, Polish Academy of Sciences, Warszawa, Poland
[2] Natural Sciences College of Science, Cardinal S. Wyszyński University, Warszawa, Poland
[3] Institute of Theoretical Physics, University of Warsaw, Warszawa, Poland

**Abstract**

We report on an extensive structural and electrical characterization of under-gate dielectric oxide insulators $Al_2O_3$ and $HfO_2$ grown by Atomic Layer Deposition (ALD). We elaborate the ALD growth window for these oxides, finding that the 40-100 nm thick layers of both oxides exhibit fine surface flatness and required amorphous structure. These layers constitute a base for further metallic gate evaporation to complete the Metal-Insulator-Semiconductor structure. Our best devices survive energizing up to ~3 MV/cm at 77 K with the leakage current staying below the state-of-the-art level of 1 nA. At these conditions the displaced charge corresponds to a change of the sheet carrier density of $3 \times 10^{13}$ $cm^{-2}$, what promises an effective modulation of the micromagnetic properties in diluted ferromagnetic semiconductors.

## 1. Introduction

One of the main aspects in the spintronic research is the development of diluted ferromagnetic semiconductors (DFS) and their use as a test bed for concept and device proofs for future applications. Ferromagnetism of these compounds is brought about by the exchange interaction among magnetic impurity spins (so far successful only for Mn) and valence band holes. Accordingly, ferromagnetism of these materials depends on the hole concentration *p* and, thus, can be controlled by the gate electric field in metal-insulator-semiconductor (MIS) structures with the magnetically doped channel. After the first successful realization of this concept [1], a greater technological progress has been achieved when high-κ dielectric oxides were employed. Except of high dielectric constants these new gate insulators offer large breakdown fields and oxides like zirconium dioxide ($ZrO_2$, $\kappa$ = 20 - 23), hafnium dioxide ($HfO_2$, $\kappa$ = 16 - 19) and aluminium oxide ($Al_2O_3$, $\kappa$ = 8 - 9) [2] are seen as the alternatives to $SiO_2$ in commercial electronics, as $HfO_2$ has already been used in 45 nm node processors [3]. On the ground of spintronic research these oxides allows now to modulate not only the Curie temperature $T_C$, but all the micromagnetic characteristics in an unprecedented wide range [4-5].

[*] sgieral@ifpan.edu.pl

These materials are grown by the Atomic Layer Deposition (ALD) method [6], which is a a self-limiting and sequential growth process, which enables using very reactive precursors. This method in general offers high conformability (films of uniform thickness can be obtained on substrates with complex shape [7]), smooth surfaces, a precise control of composition and thickness, which together with excellent patterning properties down to nanometer scale, makes it ideal solution for various under-gate dielectrics deposition. Simultaneously, all of this can be obtained at relatively low growth temperatures, reaching even below 100º C, as recently shown for ZnO [8]. This in turn makes this method ideal for low temperature budget applications like hybrid organic/inorganic and 3D electronics [9-11] or in connection with DFS like (Ga,Mn)As [12-13].

## 2. Experimental

The high-$\kappa$ thin films $HfO_2$ and $Al_2O_3$ were deposited on glass, n-Si, p-GaAs and GaMnAs substrates by the ALD method in a Savannah-100 reactor from Cambridge NanoTech Company. We used deionized water as an oxygen precursor, tetrakis(dimethylamido)hafnium (TDMAH) as a hafnium precursor and trimethylaluminum (TMA) as an aluminum precursor. The dielectric oxides were obtained at low temperature (between 80°C – 240°C) by double-exchange chemical reactions:

$$Hf[(CH_3)_2N]_4 + 2H_2O \rightarrow HfO_2 + 4HN(CH_3)_2$$
$$2Al(CH_3)_3 + 3H_2O \rightarrow Al_2O_3 + 6CH_4.$$

The structural characterization was carried out by the X-ray diffraction (XRD) method. The surface morphology and cross-section of the films were investigated by atomic force microscopy (AFM) and scanning electron microscopy (SEM). To obtain capacitor-like structures for gating, a 3 nm Cr/50 nm Au gate electrode was evaporated on the hand-painted photo-resist mask. After the lift-off the final electrical connections to the Au gate and base semiconductor were done by silver-based conductive apoxy and indium, respectively. In order to put our gating technology to the ultimate test, the area of our capacitors was about $2 \times 2$ mm$^2$.

## 3. Results and discussion

In order to increase efficiency and yield of the gated structures we started our project from elaboration of the basic growth conditions that may influence the electrical properties of gate dielectrics. For this purpose we used n-Si substrates. We began with the growth temperature ($T_g$) as it is one of the most important parameter in the ALD process, as it controls the surface coverage. We started from elaborating the ALD growth window, which is the temperature range where the growth rate is constant and the growth process is the closest to one monolayer per cycle. We established that the ALD window for HfO$_2$ is in the range between 130°C to 140°C giving the growth rate of 1.4 Å per cycle (Fig. 1a) and between 180°C to 200°C for Al$_2$O$_3$ giving the growth rate of 1.0 Å per cycle (Fig. 1b).

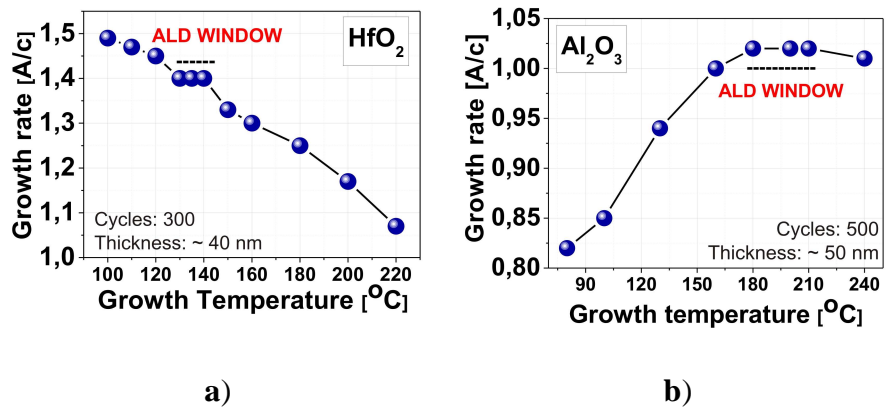

a)                                  b)

**Fig. 1**: The growth rate of the (**a**) HfO$_2$ and (**b**) Al$_2$O$_3$ films as a function of deposition temperature. ALD window is in the range of temperatures: (**a**) 130°C-140°C and (**b**) 180°C-200°C.

The XRD measurements reveal that the crystallographic structure depends on the thickness of the layer. Although Al$_2$O$_3$ grown at the ALD window conditions ($T_g$ = 200°C, Fig. 2a) shows consistently amorphous structure for the wide range of the thickness of the layers, the crystallographic structure of HfO$_2$ ($T_g$ = 135°C, Fig 2b) changes from the amorphous for the lower thicknesses to a homogenously nanocrystalline one for the very thick layers. We note that films with a thickness of above about 200 nm show a tendency to crystallize even at low process temperature. These layers crystallize in monoclinic structure. Since the amorphous structure is expected to have substantially reduced leakage current, usually associated with presence of the grain boundaries of highly crystalline films [14], this is the amorphous structure that is preferred for the gate applications. Therefore, the crystallographic dependence on the thickness of the HfO$_2$ layers instructs us that we should work with rather thin HfO$_2$ layers in order to increase technological yield of manufactured gated structures.

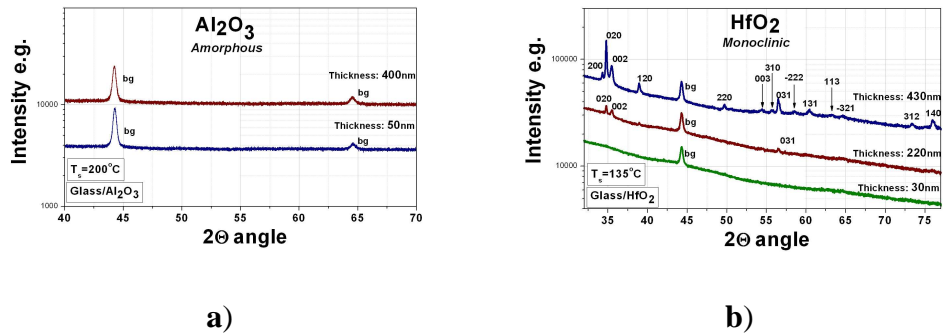

**Fig. 2**: The XRD spectra of (**a**) HfO$_2$ films grown at temperature of 135°C, and (**b**) Al$_2$O$_3$ films grown at temperature of 200°C by the ALD on glass substrates.

The leakage current, and so the gate efficiency, depends also on the surface roughness, which should be kept at the lowest possible value. We assess this parameter by AFM studies, finding that the root mean square (RMS) value of the surface roughness depends both on $T_g$ and the layer thickness. For the medium-κ Al$_2$O$_3$ the RMS decreases about twice when we reduce $T_g$ or layer thickness (see Fig. 3a and b). All these values (about 0.5 nm) are regarded as very low indicating that for gating application this oxide can be grown at its growth window conditions. The situation is completely different for HfO$_2$, because the layer grown at growth window conditions exhibits prohibitively high values of surface RMS, 4 - 16nm (see Fig. 3 d). However, we find that RMS improves substantially with reducing both $T_g$ and the layer thickness, reaching an acceptable value of 0.6 nm for $T_g$ = 85°C and thickness of 100 nm (Fig. 3 c). We select these conditions as the basic growth parameters for deposition of gate insulating layers.

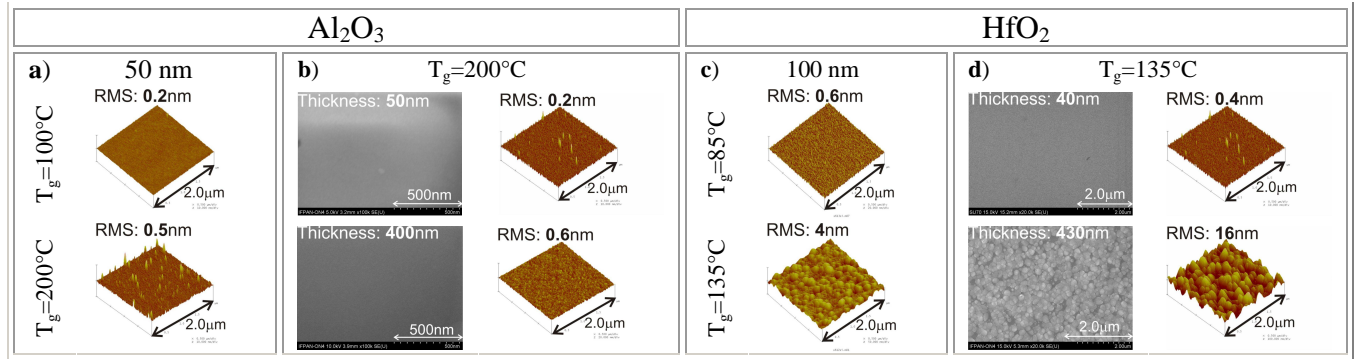

**Fig. 3**: SEM and AFM surface morphology studies of $Al_2O_3$ (**a**), (**b**) and $HfO_2$ (**c**), (**d**) thin films on Si substrate.

The ability of electric field control of magnetic anisotropy, as recently demonstrated in (Ga,Mn)As [3], has became recently a subject of intensive studies as it leads to all-electric magnetization reversal. Gated DFS or diluted magnetic semiconductors in general offers also a deep insight into the most fundamental questions of the nature of magnetic coupling in magnetically diluted systems. Therefore we concentrate our efforts on (Ga,Mn)As layers as our semiconductor channels of the prime interest. Having magnetic measurements in mind (where the experimental abilities are determined by the sheer number of spins, which for a given thickness depends on the area of the structure), but also to put our structures to the ultimate tests, we make our capacitors of a large area, typically of about 4 $mm^2$. The general layout of the capacitor MIS structure is presented in the inset to Fig. 4, whereas the typical cross-sectional SEM image constitute the main body of the figure. We find that both interfaces: semiconductor/dielectric-oxide, and the oxide/metallic-gate are fine and smooth. Moreover, a multiple SEM imaging (not shown) confirms the same structure morphology along the millimeters long sections of our devices.

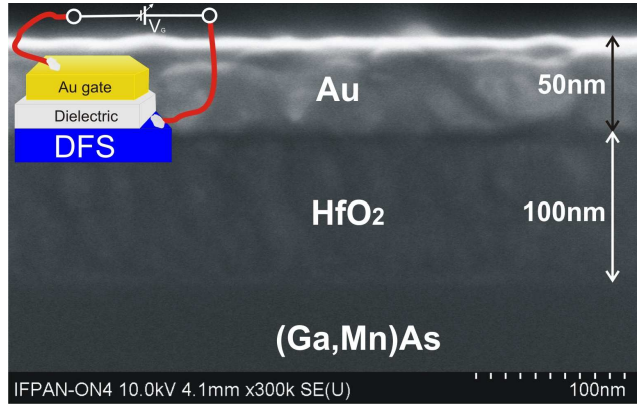

**Fig. 4**: The cross-section SEM image of the real test structure produced from Au gate (50nm) and HfO$_2$ high-κ dielectric (100nm) grown at 85°C on GaMnAs substrate. The inset shows the general layout of our capacitor structure.

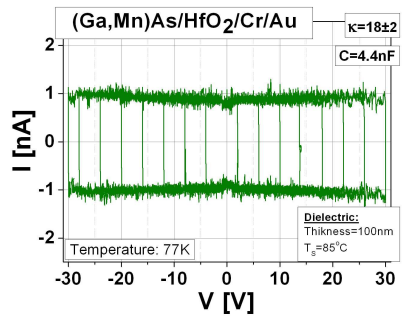

**Fig. 5**: The I-V characteristic for the (Ga,Mn)As/100 nm HfO$_2$/(3 nm Cr + 50 nm Au) capacitor recorded at 77 K. The voltage across the structure was cyclically ramped to progressively higher voltage in 4 V increments. The test was stopped at 30 V what corresponded to electric field of 3 MV/cm.

We usually define up to 12 individual capacitors in one process, which are tested for the break-down electric field in a home made I-V tester. Figure 5 presents I-V dependency for one of the best structures obtained so far on (Ga,Mn)As. Horizontalness of the I-V curve indicates that the leakage current stays below 0.1 nA for this device (or $3\times10^{-9}$ A/cm$^2$ in absolute units) up to electric field of = 3 MV/cm. Although the latter number does not compare very favorably to the best numbers reported in the literature [2,12,13], the leakage current limit puts our structures among the best contemporary state-of-the-art devices. Direct measurements of the capacitances of our structures gives dielectric constants of HfO$_2$ κ ≅ 18±2. A typical device yield for parameters given in Fig. 4 exceeds 50%.

The values presented here allow to gate in or our of the base semiconductor a charge that corresponds to surface density up to about $\Delta p_s = 3 \times 10^{13}$ cm$^{-2}$. This value already promises an effective modulation of the micromagnetic properties in DFS-based structures.

**4. Conclusions**

High quality films of high-k dielectric oxides are deposited at low temperature. Growth parameters are optimized for the monolayer growth mode and for maximum smoothness required for gating.

First working metal-insulator-semiconductor (MIS) structures based on (Ga,Mn)As are successfully fabricated exhibiting the state-of-the-art electrical parameters, offering an ability to effective gating of the magnetic properties of this DFS compound.

**Acknowledgements:**

The authors are indebted to J. Sadowski for proving (Ga,Mn)As layers for testing purposes. The work was supported in part by the European Research Council through the FunDMS Advanced Grant within the "Ideas" 7th Framework Programme of the EC, EC Network SemiSpinNet (PITN-GA-2008-215368), and InTechFun (POIG.01.03.01-00-159/08) from EU.